\documentclass[12pt,twoside]{article} 
\usepackage{psfig}
\usepackage{epsfig}
\usepackage{amssymb,wasysym,bbm,rotating,subfigure,here,citesort,bm}
\usepackage{graphicx}
\newcommand{\be}{\begin{equation}}
\newcommand{\ee}{\end{equation}}
\newcommand{\bea}{\begin{eqnarray}}
\newcommand{\eea}{\end{eqnarray}}
\newcommand{\benn}{\begin{displaymath}}
\newcommand{\eenn}{\end{displaymath}}
\newcommand{\beann}{\begin{eqnarray*}}
\newcommand{\eeann}{\end{eqnarray*}}
\newcommand\eref[1]{Eq.~(\ref{#1})}

\def\a{\alpha}
\def\d{\delta}

\def\r{\rho} 

\def\c{\raisebox{0.5mm}{$\chi$}} 
\def\bu{{\bm{u}}} 
\def\bv{{\bm{v}}} 
\def\bw{{\bm{w}}} 
\def\bx{{\bm{x}}}
\def\by{{\bm{y}}} 
\def\bz{{\bm{z}}} 
\def\fx{{\!\!\!\!}} 

\newcommand{\gtsim}{\lower-0.45ex\hbox{$>$}\kern-0.77em\lower0.55ex\hbox{$\sim$}}
\newcommand{\ltsim}{\lower-0.45ex\hbox{$<$}\kern-0.77em\lower0.55ex\hbox{$\sim$}}
\newcommand{\tr}{\mbox{tr}}    
\newcommand{\befig}{\begin{figure}}
\newcommand{\efig}{\end{figure}}
\newcommand{\betab}{\begin{table}}
\newcommand{\etab}{\end{table}}

\def\del{\partial}                              

\begin{document}
%
%
%
\pagestyle{empty}
%
%
\title{ \vspace*{-1cm} {\normalsize\rightline{CU-TP-1129}}
  \vspace*{0.cm} 
{\Large {\bf On the Projectile-Target Duality of 
  the Color Glass Condensate in the Dipole Picture} 
  \footnote{This work is supported in part by the US Department of Energy.}
}}  

\date{} \maketitle
 
%

\vspace*{-2.5cm}
 
\begin{center}
 
\renewcommand{\thefootnote}{\alph{footnote}}
 
{\large
C. Marquet$^1$\,\footnote{marquet@spht.saclay.cea.fr},   
A.H. Mueller$^2$\,\footnote{arb@phys.columbia.edu (A.H.Mueller)},
A.I. Shoshi$^2$\,\footnote{shoshi@phys.columbia.edu},  
S.M.H. Wong$^2$\,\footnote{s\_wong@phys.columbia.edu} 
} 
 
 
{\it $^1$ Service de Physique Th\'eorique, CEA/Saclay, F-91191 Gif-sur-Yvette, 
     France \\ 
     $^2$ Physics Department, Columbia University, New York, NY 10027, USA 
     }

 
\end{center}
 

\begin{abstract}
  
Recently Kovner and Lublinsky proposed a set of equations which can
be viewed as dual to JIMWLK evolution. We show that these dual equations
have a natural dipole-like structure, as conjectured by Kovner and
Lublinsky. In the high energy large $N_c$ limit these evolution equations
reduce to equations previously derived in the dipole model. We also
show that the dual evolution kernel is scheme dependent, although its
action on the weight functional describing a high energy state
gives a unique result. 

  \vspace{1.cm}
 
 
\medskip


\end{abstract}

%
%
%

\pagenumbering{roman}
\pagestyle{plain}
%
%
%
\pagenumbering{arabic}
\pagestyle{plain}
%
\makeatletter
\@addtoreset{equation}{section}
\makeatother
\renewcommand{\theequation}{\thesection.\arabic{equation}}

\newpage
\section{Introduction}
\label{Sec_Introduction} 

The Balitsky-JIMWLK equation
\cite{Balitsky:1995ub+X,Jalilian-Marian:1997jx+X,Iancu:2001ad+X,%
Weigert:2000gi} are equations governing the small-x QCD evolution for
dense partonic systems. The Balitsky equations are an infinite
hierarchy of coupled equations expressing the energy dependence of the
scattering of high energy quarks and gluons (represented by Wilson
lines in the fundamental and adjoint representations, respectively) on
a target.  The JIMWLK equation is a functional Fokker-Planck
equation~\cite{Weigert:2000gi,Blaizot:2002xy} for the small-x
evolution of the target wavefunction equivalent to the Balitsky
equations.  The Kovchegov~\cite{Kovchegov:1999yj} equation is a
simplified version of the Balitsky equations where correlations are
suppressed, leading to a relatively simple nonlinear equation for the
elastic scattering amplitude.

It has recently~\cite{Iancu:2004iy,Iancu:2005nj} been realized that the
Balitsky-JIMWLK equations miss some essential ingredients in
satisfying unitarity constraints in a realistic manner.  While these
equations accurately handle the recombination of gluons when the gluon
occupation number is large they do not properly create the growth of
the occupation number starting from a dilute system.  For that reason
they are accurate, at least in a limited energy domain, when starting
with a dense wavefunction, such as that of a big nucleus, but they are
not accurate starting from a dilute system such as an elementary
dipole.

Iancu and Triantafyllopoulos~\cite{Iancu:2004iy,Iancu:2005nj}
suggested a new equation which consists of the Balitsky hierarchy
along with a stochastic term which, in a dipole language, corresponds
to dipole creation or dipole splitting. In Ref.~\cite{Mueller:2005ut}
Mueller, Shoshi and Wong cast this equation into an equation for the
JIMWLK weight function with the addition to the usual JIMWLK
Fokker-Planck term being a fourth order functional derivative. Later
on in the paper, this extension to the JIMWLK equation will be
referred to as the MSW term. The effect of this stochastic term on the 
saturation momentum and the scattering matrix at asymptotic rapidities
has been worked out in \cite{Mueller:2004se,Iancu:2004es}.  Finally,
in Ref.~\cite{Blaizot:2005vf} both the splitting and recombination
terms were written in a compact and simple form in the relevant large
$N_c$ limit for the high-energy scattering problem.  (Parts of these
results were anticipated in Ref.~\cite{Levin:2005au}.)

Kovner and Lublinsky~\cite{Kovner:2005nq,Kovner:2005uw} have suggested
a general duality between the equations for low-density and
high-density systems.  In the above mentioned large $N_c$ limit this
duality is manifest~\cite{Blaizot:2005vf} and the equations for dipole
splitting are exactly those given in
Ref.~\cite{Iancu:2005nj,Mueller:2005ut}.  The general form of the
duality suggested in Refs.~\cite{Kovner:2005nq,Kovner:2005uw} is
likely correct and there is an ongoing effort to make this duality
more explicit~\cite{Hatta:2005rn}.

In this paper, we address the relationship between the splitting term
discussed in Ref.~\cite{Iancu:2005nj,Mueller:2005ut} and the more
general splitting terms given in Ref.~\cite{Kovner:2005uw}.  This is
partly an elaboration of discussions previously given in
Refs.~\cite{Mueller:1994gb,Salam:1995zd,Iancu:2003zr} in a closely
related context.  In particular we emphasize that the splitting term
used in Refs.~\cite{Iancu:2005nj,Mueller:2005ut} actually defines what
is meant by the large-$N_c$ limit in a high-energy scattering problem.
We recall that for the high-energy scattering of two dipoles the
leading terms in $\alpha_s\,N_c\,Y$ are in fact just the BFKL parts of
the evolution.  All multi-BFKL evolutions (multi-pomeron terms) are,
strictly speaking, higher order in $N_c.$ However, it is convenient to
define a high-energy large-$N_c$ limit where one keeps $1/N_c^2$ terms
which are enhanced by a factor
$\exp[(\alpha_P-1)Y]$~\cite{Mueller:1994gb,Salam:1995zd,Iancu:2003zr}
where $Y$ is the rapidity of the scattering and $\alpha_P$ the usual
hard pomeron intercept.  This is a natural definition since it is
exactly when $\alpha_s^2\exp[(\alpha_P-1)Y]$ is the order of one that
unitarity corrections become important. In this large-$N_c$ limit
there is no difference between the dipole splitting used in
Ref.~\cite{Iancu:2005nj,Mueller:2005ut} and the more general splitting
given in Refs.~\cite{Kovner:2005nq,Kovner:2005uw}.

While from the point of view of an evolved wavefunction the dominance
of the ``large-$N_c$'' dipole splittings is manifest it is not so
clear how this occurs as one evolves the general vertex (dipole
splitting).  We examine both generally and explicitly the scattering
to two projectile dipoles on a target dipole.  At low energies the
general splitting term and the ``large-$N_c$'' splitting term are not
the same and the difference between the two contributions is of the
same size as either contribution~\cite{Kovner:2005uw}.  However, as
one evolves to large rapidity the dominant contribution, involving the
``large-$N_c$'' splitting term, behaves as
$\alpha_s^4\exp[2(\alpha_P-1)Y]$ while the difference between the
large-$N_c$ term and the general term behaves as
$\alpha_s^4\exp[(\alpha_P-1)Y]$. This is a remarkable result.  It has
a counterpart in pomeron splitting language where Braun and
Vacca~\cite{Braun:1997nu} observed that in a particular, and natural,
scheme for defining the triple pomeron
vertex~\cite{Bartels:1994jj,Bartels:2004ef} that vertex dominates over
non-triple pomeron splittings.  This dominance seems to be exactly the
same as what has been described in comparing our large-$N_c$ splitting
and the correction terms which evolve less rapidly in rapidity.

We have also found a surprising result.  If one writes the evolution
of the JIMWLK weight function as in \eref{eq:chiW}, then we find $\c$ 
to be scheme dependent while $\c\, W_Y$ is uniquely defined.  We have 
arrived at this result in an explicit way.  We take $\c$ to be given 
by \eref{eq:chi-2}, and we find that the terms involving four derivatives 
with respect to $\rho$ do not agree with the result, Eqs. (2.38) and 
(2.39), of Ref.~\cite{Kovner:2005uw}. However, we do not believe this 
to be because one of the calculations is incorrect, but rather because 
the definitions of $\c$ are different in the two cases, Eq. (2.33) 
in \cite{Kovner:2005uw} and \eref{eq:chi-2} here.  We believe both
definitions of $\c$ are correct and that when a physical quantity,
such as the right hand side of \eref{eq:chiW}, is calculated agreement 
will be reached.  We have explicitly evaluated all the terms involving
fourth order derivatives of $\rho$ on the right hand side of 
\eref{eq:chiW}, and we find the generalized dipole picture of Kovner 
and Lublinsky\cite{Kovner:2005uw} emerges. This can be viewed as an
explicit check of the more general derivation of evolution of the
generalized dipole picture of the weight function $W_Y$ given in
Sec. \ref{dipole}. (This result has also recently been found in
Ref.~\cite{Hatta:2005ia}.) 

The outline of the paper is as follows: In Sec. \ref{jim-dual} we
briefly review JIMWLK evolution and the dual evolution discussed in
Refs.~\cite{Kovner:2005nq,Kovner:2005uw}. In Sec. \ref{dipole} we show
that the dual evolution follows that of a generalized dipole picture
as previously suggested by Kovner and Lublinsky~\cite{Kovner:2005uw}.
In Sec. \ref{Sec_Eff_Ev_HES} we show that the splitting terms not
included in the large-$N_c$ picture evolve much more slowly in energy
than those included in the high-energy large-$N_c$ approximation.  In
Sec. \ref{scheme} we explicitly evaluate terms through four
derivatives in $\rho$ and discuss the scheme dependence of the
evolution kernel.

\section{The evolution equation of the color glass condensate}
\label{jim-dual}

\subsection{The Balitsky-JIMWLK equation} 

For a left moving projectile with lightcone time $x^-$ which 
experiences at the time of the collision the effect of the color 
fields $\a^a(x^-,\bm{x})$ originated from the target, the 
Balitsky-JIMWLK equation is 
\be 
\frac{\partial}{\partial Y} W_Y[\alpha] = 
  \frac{1}{2} \int_{\bm{x}, \bm{y}}\, 
  \frac{\delta}{\delta \alpha^a(\bm{x})}\, 
  \eta^{ab}(\bm{x},\bm{y})[\alpha] \,  
  \frac{\delta}{\delta \alpha^b(\bm{y})} W_Y[\alpha]\ ,
\label{JIMWLK}
\ee 
where $Y = \ln 1/x$ is the rapidity of the small-$x$ gluons,
$\int_{\bm{x}} \equiv \int d^2 \bm{x}$ ($\bm{x}$, $\bm{y}$ denote
transverse coordinates), $\alpha^a$ is the color gauge field radiated
by the color sources in the target, and the functional $W_Y[\alpha]$
is the weight function for a given field configuration. The functional
derivatives $\d/\d \a^a(\bm{x})$ here are meant to be derivatives at
the largest $x^-$ or equivalently they stand for 
\be \frac{\d}{\d \a^a(\bm{x})} \equiv 
    \lim_{x^-\rightarrow \infty} \frac{\d}{\d \a^a(x^-,\bm{x})} \;. 
\label{eq:d/da}
\ee 
The kernel
$\eta^{ab}({\bm{x}},{\bm{y}})[\a]$ is a non--linear functional of
$\alpha$, 
\be 
\eta^{a b}(\bm{x},\bm{y})[\a] = 
  \frac{1}{\pi}\int_{\bm{z}}\,
  {\cal K}({\bm{x}, \bm{y}, \bm{z}})
  \,( 1+\tilde V^\dagger_{\bm{x}} \tilde V_{\bm{y}} 
       -\tilde V^\dagger_{\bm{x}} \tilde V_{\bm{z}} 
       -\tilde V^\dagger_{\bm{z}} \tilde V_{\bm{y}})^{ab} \ , 
\label{eta-a} 
\ee
as it depends on the Wilson lines in the adjoint representation 
$\tilde V$ and  $\tilde V^\dagger$, 
\be
\tilde{V}^\dagger_{\bm{x}}[\alpha] \,=\,{\rm P}\,{\rm exp}
 \left({\rm i}g \int dx^- T^a \alpha^a(x^-,{\bm{x}}) \right)  
\label{Vdef} 
\ee
with $(T^a)_{bc} = -i f^{abc}$, where P denotes the path--ordering 
along the light-cone coordinate $x^-$ and 
\be 
{\cal K}({\bm{x}, \bm{y}, \bm{z}}) \equiv \frac{1}{(2\pi)^2}\,
   \frac{(\bm{x}-\bm{z})\cdot(\bm{y}-\bm{z})}{
     (\bm{x}-\bm{z})^2 (\bm{z}-\bm{y})^2} \ .
\label{calK}
\ee 
%

\subsection{The dual equation to Balitsky-JIMWLK}

Given the JIMWLK equation, the dual equation can be obtained 
from JIMWLK via the duality transformations 
\cite{Blaizot:2005vf,Kovner:2005uw,Kovner:2005nq,Hatta:2005rn} 
\be 
x^- \Longleftrightarrow x^+\;, \mbox{\hspace{0.5cm}}
\frac{\d}{\d \a^a(x^-,\bm{x})} \Longleftrightarrow i \r^a(x^+,\bm{x})\;, 
\mbox{\hspace{0.5cm}} 
\a^a(x^-,\bm{x}) \Longleftrightarrow -i \frac{\d}{\d \r^a(x^+,\bm{x})} 
 \;. 
\label{dual-transf}
\ee 
This transformation is a consequence of the Lorentz boost invariance
and the arbitrariness in the labelling of which of the incoming hadrons 
is the target and which is the projectile. The resulting dual equation 
to JIMWLK is 
\be 
\frac{\partial}{\partial Y} W_Y[\r] 
  = -\frac{1}{2} \int_{\bm{x},\bm{y}}\, 
  \r^a(\bm{x})\, \eta^{ab}(\bm{x},\bm{y}) [-i\,\d/\d \r] \,
  \r^b(\bm{y}) W_Y[\r] 
\label{JIMWLK-dual-0} 
\ee 
where $\r^a(\bx)$ is now the two-dimensional charge density of the 
target at the largest $x^+$, so similarly to \eref{eq:d/da} the 
following short notation is used  
\be \r^a(\bx) = \lim_{x^+\rightarrow \infty} \r^a(x^+,\bx) 
              = \lim_{x^+\rightarrow \infty} \int dx^- \,
                \r^a(x^+, x^-,\bx) \;.
\ee 
The new $\eta^{ab}[-i\d/\d \r]$ is given by 
\be 
\eta^{a b}(\bm{x},\bm{y})[-i\,\d/\d \r] = 
  \frac{1}{\pi}\int_{\bm{z}}\,
  {\cal K}({\bm{x}, \bm{y}, \bm{z}})
  \,( 1+\tilde U^\dagger_{\bm{x}} \tilde U_{\bm{y}} 
       -\tilde U^\dagger_{\bm{x}} \tilde U_{\bm{z}} 
       -\tilde U^\dagger_{\bm{z}} \tilde U_{\bm{y}})^{ab} \ , 
\label{eta-r} 
\ee
with 
\be  \tilde{U}^\dagger_{\bm{x}}  
   = \tilde{V}^\dagger_{\bm{x}}[-i\,\d/\d \r]  
 \,=\,{\rm P}\,{\rm exp}
 \left( g \int dx^+ T^a \d/\d \r^a(x^+,{\bm{x}}) \right)  \;.
\label{Udef} 
\ee Here $P$ indicates, as usual, an ordering of the integral where
terms having larger values of $x^+$ come further to the left. Whereas
in the JIMWLK equation the emphasis is on the functional variable
$\a(x^-,\bx)$, the field experienced by the projectile, the dual
equation has the emphasis on the two-dimensional density,
$\r(x^+,\bx)$, of the target.

\section{The color dipole version of the dual equation to JIMWLK}
\label{dipole} 

In order to arrive at a color dipole version of \eref{JIMWLK-dual-0} 
upon which all our further discussions will be based, we examine 
two cases of the $W_Y$ for a single dipole and for two dipoles 
respectively to all orders in derivatives in $\r$. At the end we
make a brief remark on the case of an arbitrary number of dipoles. 

\subsection{The case of a single dipole} 
\label{single-dipole}

We denote a single dipole as 
\be W^{(0)}_Y(\bx,\by) =\frac{1}{N_c} \tr (U^\dagger_{\bm{x}} U_{\bm{y}})
                        \, \d[\r]
                  \equiv R(\bm{x},\bm{y}) \, \d[\r] \;. 
\label{eq:dual-dipole-2}
\ee 
$U^\dagger_{\bm{x}}$ is given by an identical expression as 
$\tilde U^\dagger_{\bm{x}}$ in \eref{Udef} but in the fundamental 
representation of $SU(N_c)$. The superscript $(0)$ in 
\eref{eq:dual-dipole-2} stands for the fact that one begins with an
unevolved dipole, with $Y$-evolution being given by \eref{JIMWLK-dual}. 
Thus at first order in the evolution 
\be 
\frac{\del}{\del Y} W^{(1)}_Y (\bx,\by)  
  = -\frac{1}{2} \int_{\bu,\bv}\, 
  \r^a(\bu)\, \eta^{ab}(\bu,\bv) [-i\,\d/\d \r] \,
  \r^b(\bv) \, W^{(0)}_Y (\bx,\by) \;. 
\label{JIMWLK-dual} 
\ee 
The action of $\r^a(\bv)$ on $W^{(0)}_Y(\bx,\by)$ is 
\be  \r^b(\bm{v})\, W^{(0)}_Y(\bx,\by) 
   = \frac{g}{N_c}\, \tr (U^\dagger_{\bm{x}} U_{\bm{y}} t^b) 
     \, (\d^2(\bm{v}-\bm{y})-\d^2(\bm{v}-\bm{x})) \,\d[\r]
\label{eq:rho->W0} 
\ee 
and 
\bea \r^a (\bm{u}) \r^b(\bm{v})\, W^{(0)}_Y(\bx,\by) 
   \fx & = & \fx \frac{g^2}{N_c}\, 
     \Big (-\tr (U^\dagger_{\bm{x}} U_{\bm{y}} t^b t^a) \d^2(\bm{u}-\bm{x}) 
           +\tr (U^\dagger_{\bm{x}} U_{\bm{y}} t^a t^b) \d^2(\bm{u}-\bm{y}) 
     \Big )                                             \nonumber \\  
   \fx &   & \fx \hspace{0.5cm} \times    
     \, (\d^2(\bm{v}-\bm{y})-\d^2(\bm{v}-\bm{x})) \,\d[\r] \;. 
\label{eq:rhorho->W0}
\eea 
To use \eref{eq:rho->W0} and \eref{eq:rhorho->W0} one has first to 
bring $\r^a(\bu)$ to the right of $\eta^{ab}(\bu,\bv)$. The action 
of $\r^a(\bu)$ on $(\tilde U_\bv)^{cb}$ is 
\be  \r^a(\bu) (\tilde U_\bv)^{cb} 
   = (\tilde U_\bv)^{cb}\, \r^a(\bu) +g\,(\tilde U_\bv T^a)^{cb}\, 
     \d^2(\bu-\bv) \;.  
\ee
Using \eref{eq:rho->W0} and \eref{eq:rhorho->W0} in
\eref{JIMWLK-dual} one obtains 
\bea \frac{\del}{\del Y} W^{(1)}_Y(\bx,\by) 
 \fx &=& \fx \phantom{+} 
             \frac{g^2}{\pi N_c} \int_{\bz} {\cal K}({\bm{x},\bm{y},\bm{z}}) 
  \,\{ 1+\tilde U^\dagger_{\bm{x}} \tilde U_{\bm{y}} 
        -\tilde U^\dagger_{\bm{x}} \tilde U_{\bm{z}} 
        -\tilde U^\dagger_{\bm{z}} \tilde U_{\bm{y}}\}^{ab} \, 
                                                      \nonumber \\  
 \fx & & \fx \hspace{1.8cm} \times 
    \tr (U^\dagger_\bx U_\by t^b t^a) \,\d[\r]        \nonumber \\  
 \fx & & \fx -\frac{g^2}{2\pi N_c} \int_{\bz} {\cal K}({\bm{x},\bm{x},\bm{z}}) 
  \,\{ 2 -\tilde U^\dagger_{\bm{x}} \tilde U_{\bm{z}} 
         -\tilde U^\dagger_{\bm{z}} \tilde U_{\bm{x}}\}^{ab} \, 
    \tr (U^\dagger_\bx U_\by t^b t^a) \,\d[\r]        \nonumber \\  
  \fx & & \fx -\frac{g^2}{2\pi N_c}\int_{\bz} {\cal K}({\bm{y},\bm{y},\bm{z}}) 
  \,\{ 2 -\tilde U^\dagger_{\bm{y}} \tilde U_{\bm{z}} 
         -\tilde U^\dagger_{\bm{z}} \tilde U_{\bm{y}}\}^{ab} \, 
    \tr (U^\dagger_\bx U_\by t^b t^a) \,\d[\r]        \nonumber \\  
  \fx & & \fx -\frac{g^2}{2\pi N_c} \int_{\bz} \Big ( 
                 {\cal K}({\bm{x}, \bm{x}, \bm{z}}) 
  \,\{\tilde U^\dagger_{\bm{z}} \tilde U_{\bm{x}} T^a \}^{ab} \, 
              -  {\cal K}({\bm{y}, \bm{y}, \bm{z}}) 
  \,\{\tilde U^\dagger_{\bm{z}} \tilde U_{\bm{y}} T^a \}^{ab} 
  \Big ) \,    
                                                      \nonumber \\ 
  \fx & & \fx \hspace{1.8cm} \times 
    \tr (U^\dagger_\bx U_\by t^b) \,\d[\r] \;.        \nonumber \\  
\label{dual-kov}
\eea 
To proceed further we note that $\tilde U_\bx$ can be expressed
entirely in terms of $U_\bx$ 
\be \{\tilde U_\bx\}^{ab} = 2\,\tr (U^\dagger_\bx t^a U_\bx t^b) 
\ee 
and one can use the Fierz identity 
\be t^a_{ij} t^a_{kl} 
   = \frac{1}{2}\, \Big ( \d_{il}\,\d_{jk} -\frac{1}{N_c} \d_{ij}\,\d_{kl} 
                   \Big ) 
\label{tt-lNc}
\ee
so the product of $\tilde U$'s can be rewritten as 
\be  \{\tilde U^\dagger_{\bm{x}} \tilde U_{\bm{y}}\}^{ab} 
   = \{\tilde U^\dagger_{\bm{x}}\}^{ac} \{\tilde U_{\bm{y}}\}^{cb}
   = 2^2 \,\tr (U^\dagger_\bx t^c U_\bx t^a) 
         \,\tr (U^\dagger_\by t^c U_\by t^b) 
   = 2\,\tr (U^\dagger_\by U_\bx t^a U^\dagger_\bx U_\by t^b) \;.  
\ee 
Now repeated use of \eref{tt-lNc} allows us to work out each term
in \eref{dual-kov}. For example 
\be \{\tilde U^\dagger_{\bm{x}} \tilde U_{\bm{y}}\}^{ab} \,    
    \tr (U^\dagger_\bx U_\by t^b t^a) 
  = \frac{1}{2}\, \Big (N_c-\frac{1}{N_c} \Big ) \,
                  \tr (U^\dagger_\bx U_\by)    \;, 
\label{eq:ex1}
\ee
\bea \{\tilde U^\dagger_{\bm{x}} \tilde U_{\bm{z}}\}^{ab} \,    
     \tr (U^\dagger_\bx U_\by t^b t^a)                      
  \fx & = & \fx 
     \{\tilde U^\dagger_{\bm{z}} \tilde U_{\bm{y}}\}^{ab} \,    
     \tr (U^\dagger_\bx U_\by t^b t^a)                      \nonumber \\ 
  \fx & = & \fx 
     \frac{1}{2}\, \tr (U^\dagger_\bx U_\bz)\,\tr (U^\dagger_\bz U_\by)
    -\frac{1}{2 N_c} \, \tr (U^\dagger_\bx U_\by) \;,  
\label{eq:ex2}
\eea
\be \{\tilde U^\dagger_{\bm{z}} \tilde U_{\bm{x}}\}^{ab} \,    
    \tr (U^\dagger_\bx U_\by t^b t^a) 
  = \frac{1}{2}\, \tr (U^\dagger_\bz U_\bx) \, 
                  \tr (U^\dagger_\bx U_\bz U^\dagger_\bx U_\by)
   -\frac{1}{2N_c} \, \tr(U^\dagger_\bx U_\by)  \;. 
\label{eq:ex3}
\ee
In the last line of \eref{dual-kov} there are terms with
\bea \{\tilde U^\dagger_{\bm{z}} \tilde U_{\bm{x}} T^a\}^{ab} 
    \fx & = & \fx 
    2 \,\tr (U^\dagger_\bx U_\bz t^a U^\dagger_\bz U_\bx t^c) 
    \, \{T^a\}^{cb} 
    =-2\,\tr (U^\dagger_\bx U_\bz t^a U^\dagger_\bz U_\bx (t^b t^a-t^a t^b)) 
                                                            \nonumber \\  
    \fx & = & \fx 
     -\tr (U^\dagger_\bx U_\bz) \tr (U^\dagger_\bz U_\bx t^b) 
     +\tr (U^\dagger_\bz U_\bx) \tr (U^\dagger_\bx U_\bz t^b) \;.
\eea  
This contracting with the remaining trace factor gives 
\be \{\tilde U^\dagger_{\bm{z}} \tilde U_{\bm{x}} T^a\}^{ab} 
    \tr (U^\dagger_\bx U_\by t^b) 
   = -\frac{1}{2} \Big ( 
      \tr (U^\dagger_\bx U_\bz) \tr (U^\dagger_\bz U_\by) 
     -\tr (U^\dagger_\bz U_\bx) \tr (U^\dagger_\bx U_\bz U^\dagger_\bx U_\by ) 
    \Big ) \; . 
\ee 
Substituting everything into \eref{dual-kov} all terms accompanied
by an explicit $1/N_c$ factor such as the second term in \eref{eq:ex1}, 
\eref{eq:ex2} and \eref{eq:ex3} cancel within each $\{\}^{ab}$ 
and terms with a trace of four $U$'s also cancel leaving 
\bea \frac{\del}{\del Y} \, W^{(1)}_Y(\bx,\by) 
     \fx &=& \fx -\frac{g^2}{2\pi N_c} \int_{\bz} 
                            (-2 {\cal K}({\bm{x}, \bm{y}, \bm{z}}) 
                              + {\cal K}({\bm{x}, \bm{x}, \bm{z}}) 
                              + {\cal K}({\bm{y}, \bm{y}, \bm{z}}) )
                                                          \nonumber \\ 
     \fx & & \fx \hspace{1.5cm} \times  
        \Big \{ N_c \,\tr (U^\dagger_\bx U_\by) 
               - \tr (U^\dagger_\bz U_\by) \,\tr (U^\dagger_\bx U_\bz) 
        \Big \} \,\d[\r]                                  \nonumber \\
     \fx &=& \fx -\frac{g^2 N_c}{2\pi} 
             \int_{\bz} {\cal M}({\bm{x}, \bm{y}, \bm{z}}) \,   
             \{ R(\bx,\by) - R(\bz,\by) R(\bx,\bz) \} \,\d[\r]
                                                          \nonumber \\
\label{final-dual-kov} 
\eea  
where 
\be {\cal M}({\bm{x},\bm{y},\bm{z}}) \equiv \frac{1}{(2\pi)^2}\, 
     \frac{(\bm{x}-\bm{y})^2}{(\bm{x}-\bm{z})^2 (\bm{z}-\bm{y})^2} 
\label{calM}
\ee
is the kernel of the dipole model. \eref{final-dual-kov} is valid 
for any values of $N_c$ and is the equation reminescent of the 
dipole form of the Balitsky-Kovchegov equation that most of the 
following discussions will be based. 

\subsection{The case of two dipoles and beyond} 
\label{two-dipoles}

Now starting with two dipoles, that part of $W_Y$ is 
\be W^{(0)}_Y(\bx,\bw,\by) 
   = \frac{1}{N_c} \tr (U^\dagger_\bx U^{}_\bw) \,
     \frac{1}{N_c} \tr (U^\dagger_\bw U_\by) \, \d[\r]
   \equiv R(\bx,\bw) \, R(\bw,\by) \, \d[\r] \;. 
\label{eq:dual-two-dipoles} 
\ee 
Once again we evolve this by using \eref{JIMWLK-dual-0} or 
\eref{JIMWLK-dual} and eliminating first the $\r$'s by 
acting them on $W^{(0)}_Y$. As before there are two possibilities. 
The first possibility is only one $\r$ acts on $W^{(0)}_Y$ 
\be  \r^b(\bv) W^{(0)}_Y(\bx,\bw,\by) 
   = R(\bx,\bw) \, \Big( \r^b(\bv) W^{(0)}_Y (\bw,\by) \Big ) 
    +R(\bw,\by) \, \Big( \r^b(\bv) W^{(0)}_Y (\bx,\bw) \Big ) 
\label{eq:rho->W02} 
\ee
while the other on $\eta^{ab}$, where the 
$(\r\,W^{(0)}_Y)$'s on the right hand side are given by 
\eref{eq:rho->W0}, and the second is both $\r$'s act on $W^{(0)}_Y$ 
\bea  \r^a(\bu) \r^b(\bv) W^{(0)}_Y(\bx,\bw,\by) 
  \fx &=& \fx \phantom{+}  
     R(\bx,\bw) \, \Big(\r^a(\bu) \r^b(\bv) W^{(0)}_Y (\bw,\by) \Big ) 
       \nonumber \\     
  \fx & & \fx 
    +R(\bw,\by) \, \Big(\r^a(\bu) \r^b(\bv) W^{(0)}_Y (\bx,\bw) \Big )
       \nonumber \\     
  \fx & & \fx 
    +\Big( \r^a(\bu) R(\bx,\bw) \Big ) 
     \Big( \r^b(\bv) R(\bw,\by) \Big ) \, \d[\r]
       \nonumber \\     
  \fx & & \fx  
    +\Big( \r^b(\bv) R(\bx,\bw) \Big ) 
     \Big( \r^a(\bu) R(\bw,\by) \Big ) \, \d[\r]  \,.
\label{eq:rhorho->W02} 
\eea 
The two $(\r\,\r\,W^{(0)}_Y)$'s are given by 
\eref{eq:rhorho->W0} and the $(\r\,R)$'s are essentially the same 
as $(\r\,W^{(0)}_Y)$ in \eref{eq:rho->W0} but without the 
explicit $\d[\r]$ factor. These are of course related by 
\be \Big (\r\,R(\bx,\by) \Big )\, \d[\r] 
  = \Big (\r\,W^{(0)}_Y (\bx,\by) \Big ) \,. 
\ee
Based on what we learned in Sec. \ref{single-dipole} and 
a careful examination of the form of \eref{eq:rho->W02} and
\eref{eq:rhorho->W02}, it is easy to see that  
\bea \frac{\del}{\del Y}\, W^{(1)}_Y(\bx,\bw,\by) 
   \fx &=& \fx 
          R(\bw,\by) \,\frac{\del}{\del Y}\, W^{(1)}_Y(\bx,\bw) 
        + R(\bx,\bw) \,\frac{\del}{\del Y}\, W^{(1)}_Y(\bw,\by) 
           \nonumber \\
   \fx & & \fx \hspace{-0.2cm}
        - \frac{1}{2} \int_{\bu,\bv} \eta^{ab}(\bu,\bv) [-i\d/\d \r] 
          \Big( \r^a(\bu) R(\bx,\bw) \Big ) 
          \Big( \r^b(\bv) R(\bw,\by) \Big ) \, \d[\r] 
           \nonumber \\ 
   \fx & & \fx \hspace{-0.2cm}
        - \frac{1}{2} \int_{\bu,\bv} \eta^{ab}(\bu,\bv) [-i\d/\d \r] 
          \Big( \r^b(\bv) R(\bx,\bw) \Big ) 
          \Big( \r^a(\bu) R(\bw,\by) \Big ) \, \d[\r] \,. 
           \nonumber \\    
\label{eq:evolve-2dipoles}
\eea
The expressions in the last two lines can be worked out. The first
of these is 
\bea \fx & & \fx 
     -\frac{1}{2} \int_{\bu,\bv} \eta^{ab}(\bu,\bv) [-i\d/\d \r] 
          \Big( \r^a(\bu) R(\bx,\bw) \Big ) 
          \Big( \r^b(\bv) R(\bw,\by) \Big ) \, \d[\r]    \nonumber \\
     \fx &=& \fx 
     -\frac{g^2}{2\pi N_c^2} \Big \{ \int_z 
          {\cal K} (\bw,\by,\bz) 
          \{ 1+\tilde U^\dagger_{\bm{w}} \tilde U_{\bm{y}} 
              -\tilde U^\dagger_{\bm{w}} \tilde U_{\bm{z}} 
              -\tilde U^\dagger_{\bm{z}} \tilde U_{\bm{y}}\}^{ab} 
                                                         \nonumber \\  
    \fx & & \fx \hspace{1.85cm} 
         +{\cal K} (\bx,\bw,\bz) 
          \{ 1+\tilde U^\dagger_{\bm{x}} \tilde U_{\bm{w}} 
              -\tilde U^\dagger_{\bm{x}} \tilde U_{\bm{z}} 
              -\tilde U^\dagger_{\bm{z}} \tilde U_{\bm{w}}\}^{ab} 
                                                         \nonumber \\  
    \fx & & \fx \hspace{1.85cm} 
         -{\cal K} (\bw,\bw,\bz) 
          \{ 2-\tilde U^\dagger_{\bm{w}} \tilde U_{\bm{z}} 
              -\tilde U^\dagger_{\bm{z}} \tilde U_{\bm{w}}\}^{ab} 
                                                         \nonumber \\  
    \fx & & \fx \hspace{1.85cm} 
         -{\cal K} (\bx,\by,\bz) 
          \{ 1+\tilde U^\dagger_{\bm{x}} \tilde U_{\bm{y}} 
              -\tilde U^\dagger_{\bm{x}} \tilde U_{\bm{z}} 
              -\tilde U^\dagger_{\bm{z}} \tilde U_{\bm{y}}\}^{ab} 
                          \Big \}                        \nonumber \\  
    \fx & & \fx \hspace{1.5cm} \times 
    \tr (U^\dagger_\bx U_\bw t^a) \,
    \tr (U^\dagger_\bw U_\by t^b) \,\d[\r] \,. 
\label{eq:eta-rR^2} 
\eea 
Again with the use of the Fierz identity \eref{tt-lNc}, one can
work out the terms. For example 
\be \{\tilde U^\dagger_\bw \tilde U_\by\}^{ab} 
    \tr(U^\dagger_\bx U_\bw t^a) \tr(U^\dagger_\bw U_\by t^b)
   = \frac{1}{2} \tr(U^\dagger_\bx U_\by)
    -\frac{1}{2N_c} \tr(U^\dagger_\bx U_\bw) \tr(U^\dagger_\bw U_\by) \,,
\label{eq:eta-rR^2=trU^2} 
\ee
\bea 
    \{\tilde U^\dagger_\bw \tilde U_\bz\}^{ab} 
    \tr(U^\dagger_\bx U_\bw t^a) \tr(U^\dagger_\bw U_\by t^b)
   \fx &=& \fx \frac{1}{2} 
     \tr(U^\dagger_\bx U_\bz U^\dagger_\bw U_\by U^\dagger_\bz U_\bw)
                                                      \nonumber \\ 
   \fx & & \fx 
    -\frac{1}{2N_c} \tr(U^\dagger_\bx U_\bw) \tr(U^\dagger_\bw U_\by) \,,
\label{eq:eta-rR^2=trU^6-1} 
\eea 
\bea 
    \{\tilde U^\dagger_\bz \tilde U_\bw\}^{ab} 
    \tr(U^\dagger_\bx U_\bw t^a) \tr(U^\dagger_\bw U_\by t^b)
   \fx &=& \fx \frac{1}{2} 
     \tr(U^\dagger_\bx U_\bw U^\dagger_\bz U_\by U^\dagger_\bw U_\bz)
                                                      \nonumber \\ 
   \fx & & \fx 
    -\frac{1}{2N_c} \tr(U^\dagger_\bx U_\bw) \tr(U^\dagger_\bw U_\by) \,.
\label{eq:eta-rR^2=trU^6-2}  
\eea 
In fact if one ignores the $\cal K$ factors, all terms in 
\eref{eq:eta-rR^2} with $\{\tilde U^\dagger \tilde U_\bz\}^{ab}$ 
give the same expression on the right hand side as 
\eref{eq:eta-rR^2=trU^6-1}, all terms with 
$\{\tilde U^\dagger_\bz \tilde U\}^{ab}$ give 
\eref{eq:eta-rR^2=trU^6-2} and lastly all terms without either
$\tilde U_\bz$ or $\tilde U^\dagger_\bz$ give \eref{eq:eta-rR^2=trU^2}. 
Substituting all these back into \eref{eq:eta-rR^2}, terms  
with a $1/N_c$ factor cancel among themselves within each $\{\}^{ab}$
and one is left with 
\bea \fx & & \fx 
     -\frac{1}{2} \int_{\bu,\bv} \eta^{ab}(\bu,\bv) [-i\d/\d \r] 
          \Big( \r^a(\bu) R(\bx,\bw) \Big ) 
          \Big( \r^b(\bv) R(\bw,\by) \Big ) \, \d[\r]    \nonumber \\
     \fx &=& \fx 
     -\frac{g^2}{4\pi N_c^2} \int_z 
          \Big ( {\cal K}(\bw,\by,\bz)+{\cal K}(\bx,\bw,\bz)
                -{\cal K}(\bw,\bw,\bz)-{\cal K}(\bx,\by,\bz) \Big )
                                                         \nonumber \\
     \fx & & \fx \hspace{1.6cm} \times
     \Big ( 2 \tr(U^\dagger_\bx U_\by)
           -\tr(U^\dagger_\bx U_\bz U^\dagger_\bw U_\by U^\dagger_\bz U_\bw)
           -\tr(U^\dagger_\bx U_\bw U^\dagger_\bz U_\by U^\dagger_\bw U_\bz)
     \Big ) \, \d[\r]                                    \nonumber \\    
\eea 
Incidentally swapping $\r^a$ and $\r^b$ in the above equation gives 
exactly the same result so the last two lines of \eref{eq:evolve-2dipoles} 
are identical. Finally \eref{eq:evolve-2dipoles} becomes 
\bea \fx & & \fx \hspace{-0.8cm} 
      \frac{\del}{\del Y}\, W^{(1)}_Y(\bx,\bw,\by)
                                                         \nonumber \\    
     \fx &=&  
          R(\bw,\by) \,\frac{\del}{\del Y}\, W^{(1)}_Y(\bx,\bw) 
        + R(\bx,\bw) \,\frac{\del}{\del Y}\, W^{(1)}_Y(\bw,\by) 
                                                         \nonumber \\
    \fx & & \fx \hspace{0.0cm} 
     -\frac{g^2}{2\pi N_c^2} \int_z 
          \Big ( {\cal K}(\bw,\by,\bz)+{\cal K}(\bx,\bw,\bz)
                -{\cal K}(\bw,\bw,\bz)-{\cal K}(\bx,\by,\bz) \Big )
                                                         \nonumber \\
     \fx & & \fx \hspace{1.4cm} \times
     \Big ( 2 \tr(U^\dagger_\bx U_\by)
           -\tr(U^\dagger_\bx U_\bz U^\dagger_\bw U_\by U^\dagger_\bz U_\bw)
           -\tr(U^\dagger_\bx U_\bw U^\dagger_\bz U_\by U^\dagger_\bw U_\bz)
     \Big ) \, \d[\r]                                    \nonumber \\    
    \fx &=& 
          R(\bw,\by) \,\frac{\del}{\del Y}\, W^{(1)}_Y(\bx,\bw) 
        + R(\bx,\bw) \,\frac{\del}{\del Y}\, W^{(1)}_Y(\bw,\by) 
           \nonumber \\
    \fx & & \fx \hspace{0.0cm} 
     -\frac{g^2}{2\pi N_c} \int_z 
          \Big ( {\cal M}(\bx,\by,\bz) -{\cal M}(\bw,\by,\bz)
                -{\cal M}(\bx,\bw,\bz) \Big )            \nonumber \\ 
    \fx & & \fx \hspace{1.4cm} \times
     \Big ( 2 R(\bx,\by) 
           -\frac{1}{N_c} [
            \tr(U^\dagger_\bx U_\bz U^\dagger_\bw U_\by U^\dagger_\bz U_\bw)
           +\tr(U^\dagger_\bx U_\bw U^\dagger_\bz U_\by U^\dagger_\bw U_\bz)]
     \Big ) \, \d[\r]                                    \nonumber \\    
\eea
Apart from the gauge field $A_\mu$'s in the $U$'s and $U^\dagger$'s 
have been replaced by $\d/\d\r$ here, the presence of $\d[\r]$ and 
a prefactor, this last expression is the same equation as Eq. (121) 
of the first paper in Ref.~\cite{Balitsky:1995ub+X}. Using 
\eref{final-dual-kov} one can see that the terms in the second line 
is down by a factor of $1/N^2_c$ in comparison to those in the 
first line. Therefore in the large $N_c$ limit, the evolution equation
for two dipoles simplifies to a ``product rule'' like form  
\be \frac{\del}{\del Y}\, W^{(1)}_Y(\bx,\bw,\by) \simeq 
      R(\bw,\by) \,\frac{\del}{\del Y}\, W^{(1)}_Y(\bx,\bw) 
    + R(\bx,\bw) \,\frac{\del}{\del Y}\, W^{(1)}_Y(\bw,\by) \;. 
\ee
Because there are only two $\r$'s in \eref{JIMWLK-dual-0} and they
can act on a maximum of two different $R$'s, this product rule  
holds for the evolution of any arbitrary number of dipoles in 
the large $N_c$ limit. One can conclude that in this limit each 
dipole evolves independently of each other.

\section{Dipole model versus dual evolution in the low gluon density 
region} 
\label{Sec_Eff_Ev_HES}
In this section we consider the scattering of two elementary dipoles
off an evolved target.  The latter is considered to be dilute (low
gluon density) so that gluon mergings can be neglected. We construct
the wavefunction of an evolved target starting with an elementary
dipole by using the dipole model~\cite{Mueller:1993rr,Mueller:1994jq}
and the dual evolution
equation~\cite{Kovner:2005uw,Kovner:2005nq,Hatta:2005rn}.  With the
evolution operator recently worked out by Mueller, Shoshi and Wong
(MSW)~\cite{Mueller:2005ut} to extend the JIMWLK equation to the low
gluon density region, we obtain a scattering amplitude which in the
dipole model is dominated by the scattering of the two projectile
dipoles off {\it two different} dipoles in the evolved target. Kovner
and Lublinsky~\cite{Kovner:2005uw} have shown that the dual evolution
equation gives additional terms, missed by the MSW evolution operator,
which allow for the two projectile dipoles also to scatter off {\it a
single} dipole in the evolved target.  In this section we show that
the multiple interaction of the target dipole is suppressed as compared
to the single interaction of different target dipoles at high
rapidities. The two contributions are comparable only at very low
rapidities where the target contains only few gluons. At high
rapidities with a large number of gluons in the target the effective
operator of Ref.~\cite{Mueller:2005ut} gives correctly the leading
contribution to the scattering amplitude.
\subsection{Low gluon density evolution in the dipole model}
Let us consider the evolution of an unevolved single dipole in the
weak field limit 
\be 
W_{Y\,(2)}^{(0)}({\bm x}, {\bm y}) =\frac{1}{4N_c}
(\delta^b_{\bm{x}}-\delta^b_{\bm{y}})^2 \delta[\rho] \equiv
R_{(2)}^{(0)}(\bx,\by) \,\delta[\rho]\ . 
\ee
Here the subscript $(2)$ stands for the second order derivative 
in $\rho$ and the superscript $(0)$ denotes the unevolved state
of the dipole that one begins with. The $\d^b_\bx$ here is given by  
\be \d_{\bx}^b = g \int_{-\infty}^{\infty} dx^+ 
    \frac{\d}{\d\rho^b(x^+,\bx)} \,.  
\label{eq:delta-def} 
\ee
The first step in the evolution in the extended JIMWLK equation as 
given by the dipole model reads~\cite{Mueller:2005ut}
\be 
\frac{\del}{\del Y}\ W^{(1)}_Y \,= 
(\c_2 + \c_4^{\mbox{\tiny{MSW}}})\ W_{Y\,(2)}^{(0)} \ ,
\label{Eq_Ev_MSW}
\ee
where the operator $\c_2$ gives the BFKL evolution and
$\c_4^{\mbox{\tiny{MSW}}}$ splits one scattering dipole into two
scattering dipoles at every evolution step. In the JIMWLK language, in
terms of fields, these have been calculated in
Ref.~\cite{Mueller:2005ut}.  Using the relation between the fields and
charge densities $-\nabla_{\perp}^2\,\alpha^a(x^+, {\bm x}) =
\rho^a(x^+, {\bm x})$, they can be expressed in terms of charge
densities
\be 
\c_2 =  \frac{1}{2g^2N_c}\,\frac{g^2 N_c}{2 \pi}
        \int_{\bu,\bv,\bz} {\cal M}(\bu,\bv,\bz)\ 
        \left[(\d^c_\bu-\d^c_\bz)\,(\d^d_\bz-\d^d_\bv)\right] 
        \left( T^c T^d \right)_{ab} \
        \rho^a(\bu) \rho^b(\bv) 
\ee 
and 
\be
\c_4^{\mbox{\tiny{MSW}}} = -\frac{1}{16 g^2 N_c^3}\,
   \frac{g^2 N_c}{2 \pi}
   \int_{\bu,\bv,\bz} {\cal M}(\bu,\bv,\bz)\ 
   \left[(\d^c_{\bu}-\d^c_{\bz})^2\right]\, 
   \left[(\d^d_{\bz}-\d^d_{\bv})^2\right] 
   \rho^a(\bu) \rho^a(\bv)  \,.  
\label{Eq_chi_MSW}
\ee

\begin{figure}[!ht]
\setlength{\unitlength}{1.cm}
\begin{center}
\epsfig{file=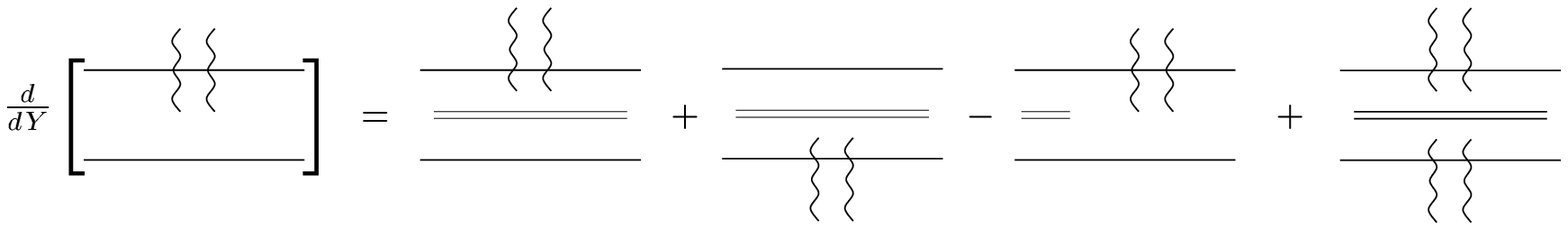, width=14.8cm}
\end{center}
\caption{One step of evolution of an unevolved single dipole in the
  dipole model: The first three graphs on the r.h.s. represent 
  the BFKL evolution of the original dipole and the last graph the
  splitting of the original dipole into two new dipoles which
  simultaneously scatter off the projectile dipoles.} 
\label{Fig_MSW_1S}
\end{figure}

Acting with these operators on the $W_{Y\,(2)}^{(0)}(\bx, \by)$ 
in Eq.~(\ref{Eq_Ev_MSW}), one obtains 
\bea
\frac{\del}{\del Y}\ W^{(1)}_Y ({\bm x}, {\bm y})\!\!&=&\!\!\! 
  \frac{g^2 N_c}{2 \pi} \int_{{\bm{z}}} 
  {\cal M}({\bm{x}},{\bm{y}},{\bm{z}})
\label{Eq_MSW_1s} 
\\
&\times& \!\!\!\!\!
\left[ R_{(2)}^{(0)}({\bm x}, {\bm z}) + R_{(2)}^{(0)}({\bm z}, {\bm
    y}) - R_{(2)}^{(0)}({\bm x}, {\bm y}) + 
R_{(2)}^{(0)}({\bm x}, {\bm z})R_{(2)}^{(0)}({\bm z}, {\bm y}) \right]
\delta[\rho] \ ,
\nonumber 
\eea 
where the first three terms on the right-hand side (r.h.s.) represent the
BFKL evolution of the original dipole ($\bx$, $\by$) while the fourth
term describes the splitting of the original dipole into two new
dipoles (${\bm x}$, ${\bm z}$) and (${\bm z}$, ${\bm y}$) which
simultaneously scatter off the projectile. This first step 
in the evolution of an unevolved single dipole is shown in
Fig.~\ref{Fig_MSW_1S}.

\begin{figure}[!ht]
\setlength{\unitlength}{1.cm}
\begin{center}
\epsfig{file=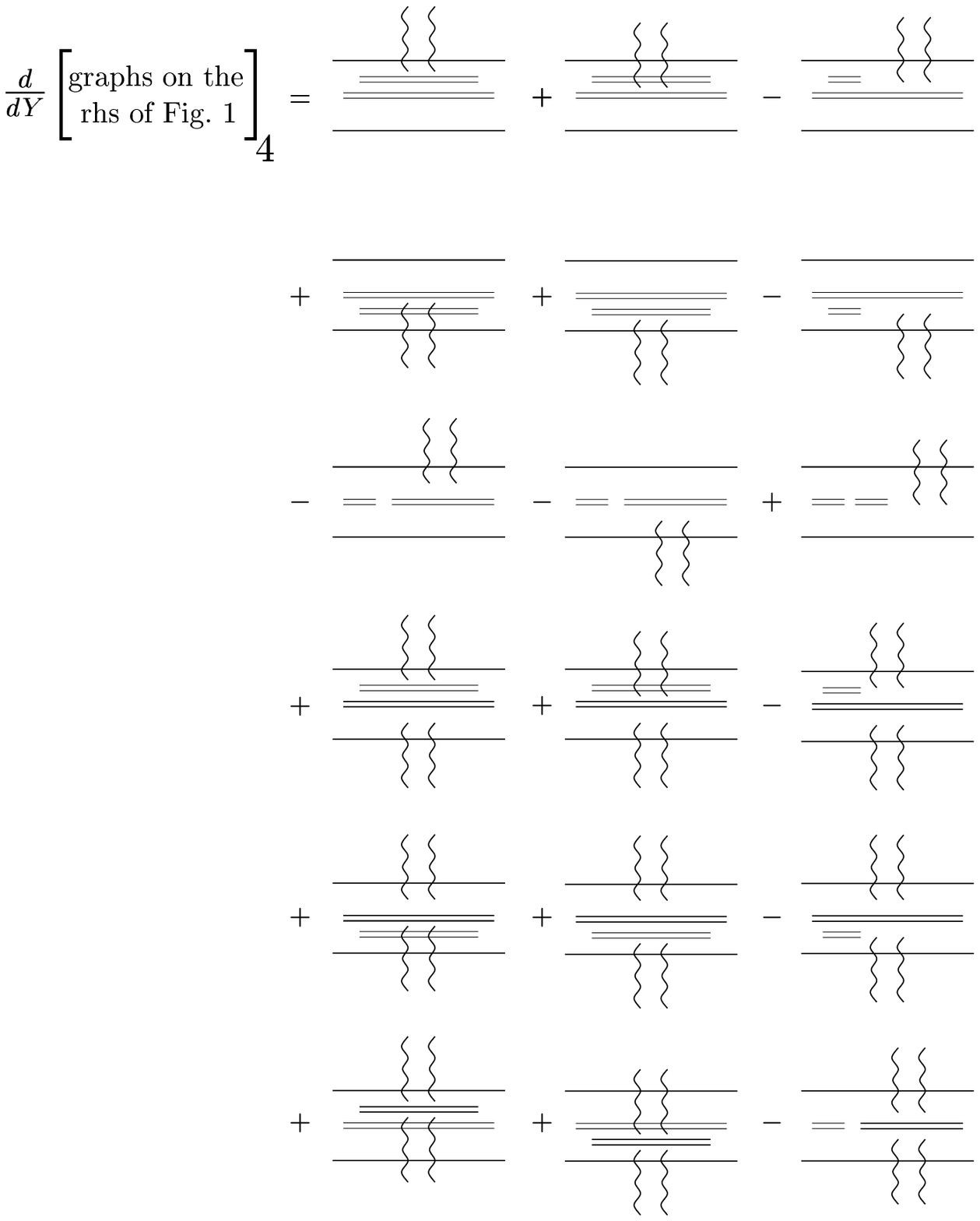, width=13cm}
\end{center}
\caption{One step of evolution of the graphs on the 
  r.h.s. of Fig.~\ref{Fig_MSW_1S} in the dipole model. More than four 
  gluon interactions with the projectile dipoles are excluded.}
\label{Fig_MSW_2S}
\end{figure}

Let us now consider the second step of the evolution.  A graphical 
representation of the result is shown in Fig.~\ref{Fig_MSW_2S}. 
This is simply obtained by evolving one step further the graphs on 
the r.h.s. of Fig.~\ref{Fig_MSW_1S} according to the dipole model in the 
extended evolution, i.e., by acting on them with $\c_2$
and $\c_4^{\mbox{\tiny{MSW}}}$ as in the first step. In 
Fig.~\ref{Fig_MSW_2S} the first five lines come from the BFKL
evolution of the four graphs in Fig.~\ref{Fig_MSW_1S},
$\c_2\,(W^{(1)}_{Y\,(2)}+W^{(1),\mbox{\tiny{MSW}}}_{Y\,(4)})$ with 
\be W^{(1),\mbox{\tiny{MSW}}}_{Y\,(4)}({\bm x}, {\bm y}, {\bm z}) 
    \equiv R_{(2)}^{(0)}({\bm x}, {\bm z})R_{(2)}^{(0)}({\bm z}, {\bm y})\ 
\delta[\rho] \;. 
\ee 
The last line results by acting with 
$\c_4^{\mbox{\tiny{MSW}}}$ on the first three graphs in  
Fig.~\ref{Fig_MSW_1S}, \linebreak
$\c_4^{\mbox{\tiny{MSW}}} W^{(1)}_{Y\,(2)}$.
The subscript $[\ ]_4$ in Fig.~\ref{Fig_MSW_2S} and
Eq.~(\ref{Eq_Ev_A}) indicates that we have dropped in the wavefunction
terms having more than four orders in $\rho$-derivatives, like
$\c_4^{\mbox{\tiny{MSW}}} W^{(0),\mbox{\tiny{MSW}}}_{Y\,(4)}$, since
for our purpose it is enough to consider the scattering of the evolved
target on only two projectile dipoles.

\begin{figure}[!ht]
\setlength{\unitlength}{1.cm}
\begin{center}
\epsfig{file=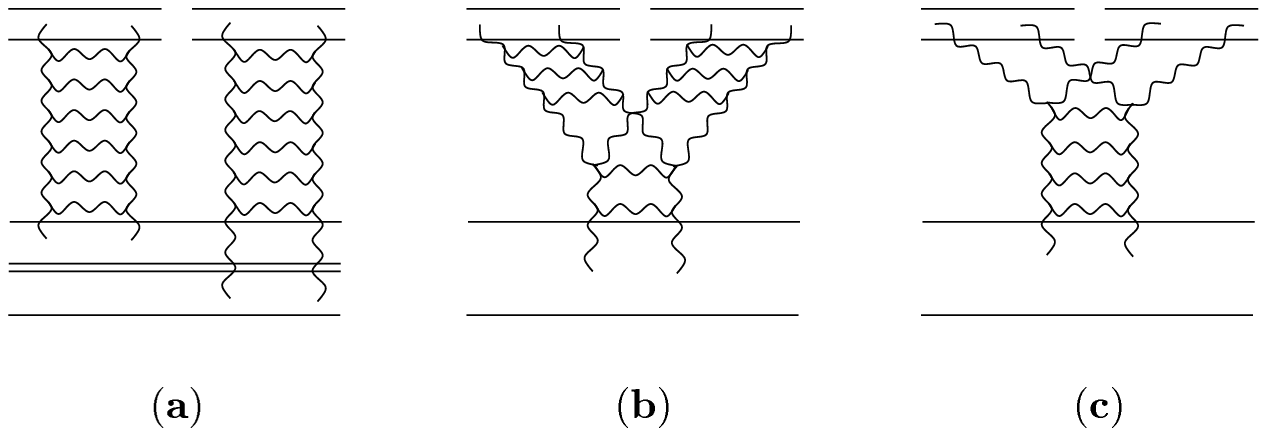, width=13cm}
\end{center}
\caption{An evolved target scattering off two elementary dipoles in
  the Pomeron language: (a) double Pomeron interaction, (b) a partly
  evolved Pomeron splitts into two Pomerons which after some evolution
  scatter off the two dipoles, (c) a fully evolved Pomeron
  scatters at the end of its evolution off the two dipoles.}
\label{Fig_Ev_T_2D}
\end{figure}

In formula the second step of the evolution is 
\be
\frac{\del}{\del Y} \left[W^{(2)}_Y\right]_4 = 
\c_2\, W^{(1)}_{Y\,(2)} + 
\c_2\, W^{(1),\mbox{\tiny{MSW}}}_{Y\,(4)} + 
\c_4^{\mbox{\tiny{MSW}}}\, W^{(1)}_{Y\,(2)} \ . 
\label{Eq_Ev_A}
\ee 
It is easy to check that this evolution equation holds for any step
of the evolution. Summing over all dipoles numbers
(superscript), the evolved wavefunction which contains
up to four $\rho$-derivatives reads
\be
\frac{\del}{\del Y} \left[W_Y\right]_4 
  = \frac{d}{dY} W_{Y\,(2)} 
   +\frac{\del}{\del Y} W^{\mbox{\tiny{MSW}}}_{Y\,(4)} 
\ee
with 
\bea
\frac{\del}{\del Y} W_{Y\,(2)} &=& \c_2\ W_{Y\,(2)} \ ,
\label{Eq_BFKL_Ev}\\
\frac{\del}{\del Y} W^{\mbox{\tiny{MSW}}}_{Y\,(4)} 
  &=& \c_2\ W^{\mbox{\tiny{MSW}}}_{Y\,(4)} 
     +\c_4^{\mbox{\tiny{MSW}}}\ W_{Y\,(2)} \ . 
\label{Eq_MSW_W_Ev}
\eea
In Eq.~(\ref{Eq_MSW_W_Ev}) $\c_4^{\mbox{\tiny{MSW}}}\ W_{Y\,(2)}$
converts one into two simultaneously interacting target dipoles at any
step of evolution. Both dipoles evolve in the subsequent steps of 
the evolution according to the BFKL equation, given by $\c_2\ 
W^{\mbox{\tiny{MSW}}}_{Y\,(4)}$, before they scatter on the two
projectile dipoles. If the two dipoles are created in the first step
of the evolution (fourth graph on the rhs of Fig.~\ref{Fig_MSW_1S}), 
then the subsequent BFKL evolution of both dipoles gives for the 
rapidity dependence of the wavefunction 
\be 
W^{\mbox{\tiny{MSW}}}_{Y\,(4)}\ \sim \ e^{2\,(\alpha_P-1)\,Y}\ 
W^{\mbox{\tiny{MSW}}}_{Y=0\,(4)} 
\label{Eq_MSW_ED}
\ee
with $\alpha_P - 1 = (4 \alpha_s N_c \ln 2)/\pi$. When considering the
scattering of the evolved target dipole off the two projectile dipoles,
(\ref{Eq_MSW_ED}) gives for the $T$-matrix 
\be
T^{\mbox{\tiny{\,2\,Pom}}}(Y) \ \sim \ \alpha^4_s \ e^{2\,(\alpha_P-1)\,Y} \ .
\label{Eq_T_DM}
\ee
In the Pomeron language the interaction of the evolved target dipole
with the two projectile dipoles occurs via the double Pomeron exchange
as shown in Fig.~\ref{Fig_Ev_T_2D}(a). However, if the two target
dipoles are generated after several previous steps of BFKL evolution 
of a single dipole (last line in Fig.~\ref{Fig_MSW_2S}), then the
rapidity dependence is partly that of a single Pomeron and partly that
of a double Pomeron exchange, $e^{(\alpha_P-1)\,y}
e^{2\,(\alpha_P-1)\,(Y-y)}$, which is suppressed as compared with the
pure double Pomeron exchange. The later case is shown in
Fig.~\ref{Fig_Ev_T_2D}(b).

\subsection{Low gluon density evolution in the dual equation} 
\begin{figure}[!ht]
  \setlength{\unitlength}{1.cm}
\begin{center}
\epsfig{file=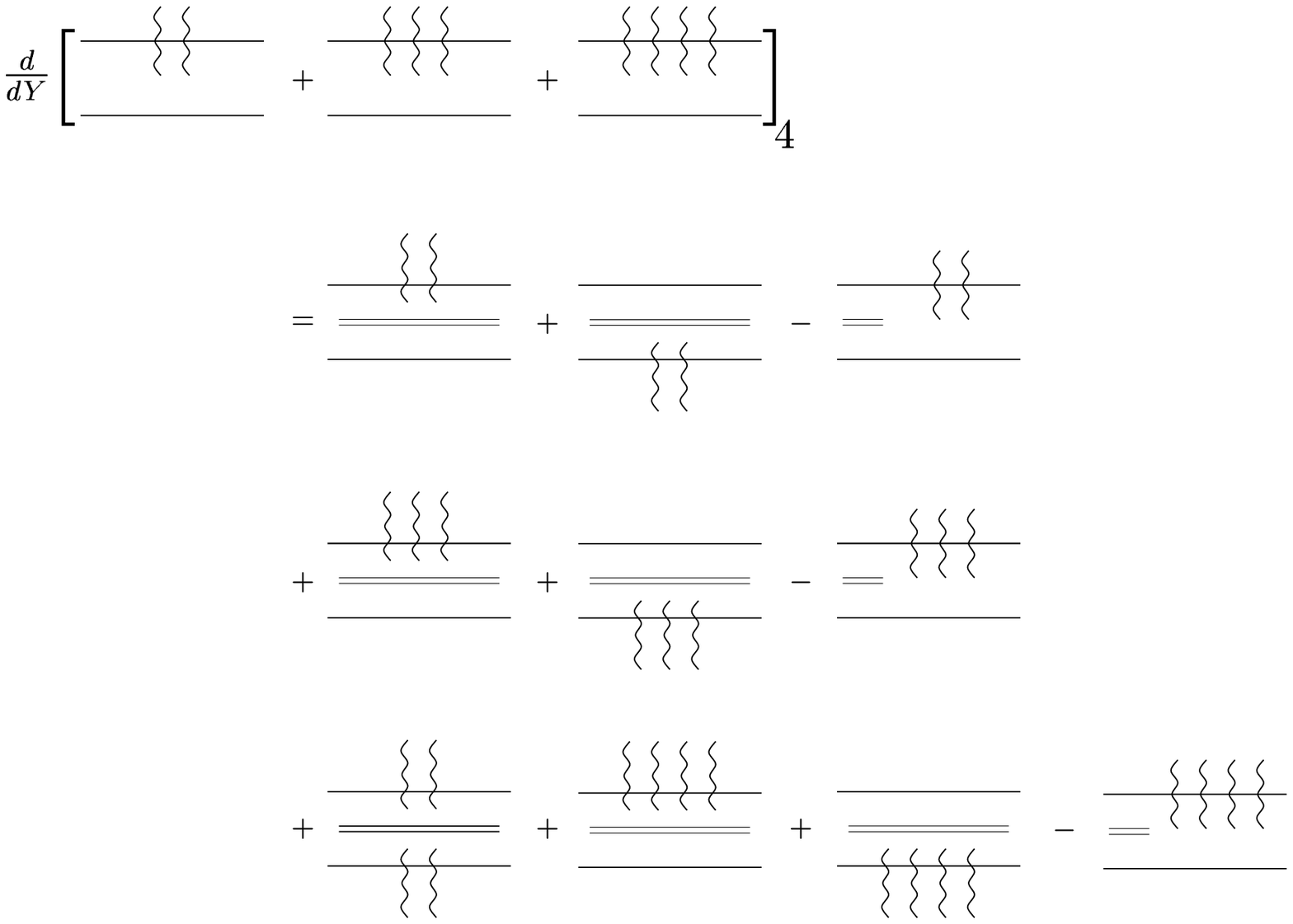, width=14cm}
\end{center}
\caption{One step of evolution of an unevolved single dipole according
  to the dual equation up to the fourth order in $\rho$-derivatives.}
\label{Fig_KL_1S}
\end{figure}

The first step of evolution of an unevolved single dipole as given 
by the dual equation \eref{JIMWLK-dual-0} up to the fourth order in 
$\rho$-derivatives (see \eref{JIMWLK-dual} and \eref{final-dual-kov}) is 
\bea 
\frac{\del}{\del Y} \Big [W^{(1)\,\mbox{\scriptsize dual}}_Y \Big ]_4 
    \fx &=& \fx  
\frac{\del}{\del Y} \Big [ W^{(1)}_{Y\,(2)} + W^{(1)}_{Y\,(3)} 
     +W^{(1)}_{Y\,(4)} \Big ]_4                
\label{Eq_KLWMIJ_1s}                                  \\ 
   \fx &=& \fx \frac{g^2 N_c}{2 \pi} 
   \int_{{\bm{x}},{\bm{y}},{\bm{z}}} {\cal M}({\bm{x}},{\bm{y}},{\bm{z}}) 
\nonumber\\ 
\fx & & \fx \times 
\Big [ \phantom{+} 
      R_{(2)}^{(0)}({\bm x}, {\bm z}) +R_{(2)}^{(0)}({\bm z}, {\bm y}) 
    - R_{(2)}^{(0)}({\bm x}, {\bm y}) 
\nonumber\\ 
\fx & & \fx \phantom{\times \Big [}
     +R_{(3)}^{(0)}({\bm x}, {\bm z}) + R_{(3)}^{(0)}({\bm z}, {\bm
    y}) - R_{(3)}^{(0)}({\bm x}, {\bm y}) 
\nonumber\\
\fx & & \fx \phantom{\times \Big [}
  +R_{(2)}^{(0)}({\bm x}, {\bm z})R_{(2)}^{(0)}({\bm z},{\bm y}) 
\nonumber \\
\fx & & \fx \phantom{\times \Big [}
  +R_{(4)}^{(0)}({\bm x}, {\bm z}) + R_{(4)}^{(0)}({\bm z}, {\bm
    y}) - R_{(4)}^{(0)}({\bm x}, {\bm y})\Big ] \delta[\rho] 
\nonumber 
\eea 
with $R_{(n)}^{(0)}\delta[\rho]$ denoting the $n$-th order
$\rho$-derivative term in the expansion of $W^{(0)}_Y$ in
\eref{eq:dual-dipole-2}. Fig. \ref{Fig_KL_1S} shows
\eref{Eq_KLWMIJ_1s} in graphical form. {\em Note that the dual
evolution allows for a single target dipole to interact also via three
and four gluon exchange while in the dipole model a single target
dipole interacts only via two gluon exchange.}

\begin{figure}[!ht]
\setlength{\unitlength}{1.cm}
\begin{center}
\epsfig{file=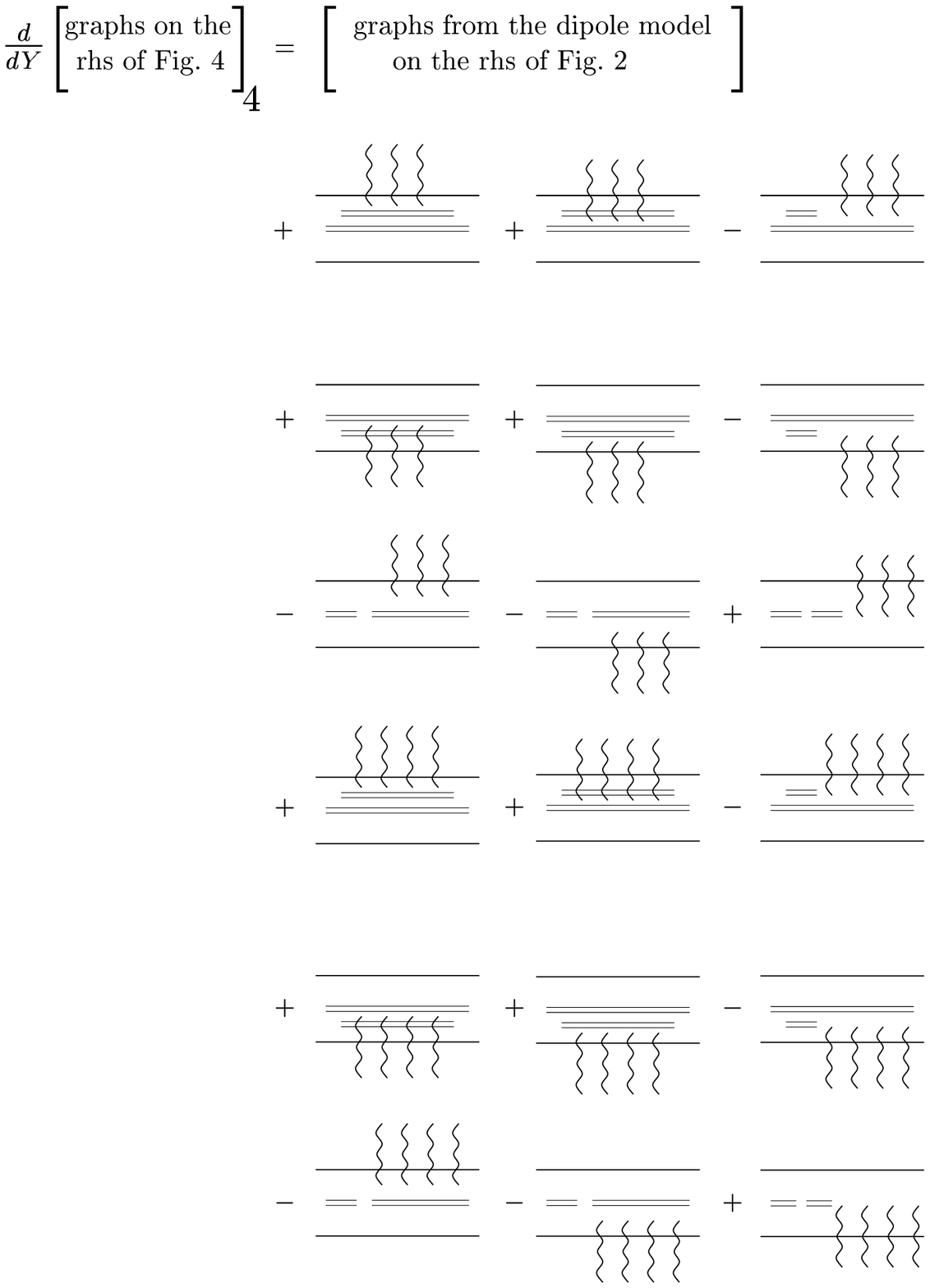, width=13cm}
\end{center}
\caption{One step of evolution of the graphs on the 
  r.h.s. of Fig.~\ref{Fig_KL_1S} according to the dual evolution up to
  the fourth order in $\rho$-derivatives.} 
\label{Fig_KL_2S}
\end{figure}

The second step of the evolution follows simply by applying the 
same rule as in the first step for the created dipoles inside 
the evolved target dipole. The resulting graphs are shown in 
Fig.~\ref{Fig_KL_2S}.  As compared with the dipole model, the extra
graphs in the dual evolution are those with three and four gluons
attached at the end of the BFKL evolution to a single target dipole.
Further steps of evolution will only add some more dipoles according 
to BFKL but will not split the three and four gluons and attach them
to different target dipoles. Thus, the rapidity dependence of the
evolved target wavefunction where three or four gluons are attached to
a single target dipole, $\overline{W}_Y$, is that of a 
single BFKL Pomeron 
\be
\overline{W}_{Y\,(i)} \sim \ e^{(\alpha_P-1)Y}\,\overline{W}_{Y=0\,(i)} 
\label{Eq_KL_4g}
\ee
where $i=3$ or $4$ stands for three or four gluons respectively. 
This leads to the following rapidity dependence of the 
$T$-matrix for the scattering of the two projectile dipoles via
two-gluon exchange with a single dipole in the evolved target
\be T^{\mbox{\tiny{\,1\,Pom$\to$4g}}}(Y) \ \sim \ \alpha^4_s 
    \ e^{\,(\alpha_P-1)\,Y} \ .
\label{Eq_T_KL}
\ee
In the Pomeron language this result is shown in
Fig.~\ref{Fig_Ev_T_2D}(c) where a single Pomeron interacts at the end
of its evolution via four gluon exchange with two dipoles.  Note that
the three gluon exchange part $\overline{W}_{Y\,(3)}$ will drop out
when considering the interactions with two projectile dipoles.

The wavefunction of an evolved target dipole at the fourth order 
of $\rho$-derivatives as given by the dual equation can be separated  
into two parts in the form 
\be 
W^{\mbox{\scriptsize dual}}_{Y\,(4)} 
   = W^{\mbox{\tiny MSW}}_{Y\,(4)} + \overline{W}_{Y\,(4)} \,,   
\label{Eq_KL_MSW}
\ee
where the first part is in the dipole model and the second is not. 
Eq.~(\ref{Eq_KL_MSW}) tells us that at high rapidities $W^{\mbox{\tiny
MSW}}_{Y\,(4)}$ gives the dominant contribution to the scattering
matrix since $W^{\mbox{\tiny MSW}}_{Y\,(4)} \gg \overline{W}_{Y\,(4)}$
(see \eref{Eq_MSW_ED} and \eref{Eq_KL_4g}). This is the same result
that was recognized already in Ref.~\cite{Braun:1997nu}: the double
Pomeron interaction (Fig.~\ref{Fig_Ev_T_2D}(a) and
Eq.~(\ref{Eq_T_DM})) dominates over the single Pomeron which interacts
at the end of its evolution with the two external dipoles
(Fig.~\ref{Fig_Ev_T_2D}(c) and Eq.~(\ref{Eq_T_KL})).

In the dipole model~\cite{Mueller:1993rr,Mueller:1994jq} our main
result reflects the fact of dipole counting in the scattering of two
external dipoles with the evolved target: when both external dipoles
can scatter off different target dipoles, each of them participates
with the factor $\alpha^2_s\ n(Y)$, thus, giving $T \sim [\alpha^2_s\ 
n(Y)]^2$, with the dipole number density $n(Y) \sim
\exp{[(\alpha_P-1)Y]}$. On the other hand, when both external dipoles
are forced to scatter off a single target dipole, the one scattering first
gives the factor $\alpha^2_s\ n(Y)$ while the second one has only one
possibility $\alpha^2_s \times 1$, leading to $T \sim [\alpha^4_s\ 
n(Y)]$. Note that only at very low rapidity where $n(Y)$ is of order
one, the two cases give similar results.

\section{Scheme dependence of the evolution kernel; explicit calculations} 
\label{scheme}

In Ref. \cite{Kovner:2005uw} the evolution in rapidity of the target 
wave function was written in the Hamiltonian form. Two versions 
of this Hamiltonian were given explicitly in terms of derivatives of 
$\r$ up to the fourth derivative: one for the general 
case and the other for when one is dealing exclusively with dipoles. 
The problem is that the derivative form of these two Hamiltonians 
did not appear to be equivalent to each other. On top of that the 
dipole version of this Hamiltonian was emphatically pointed out to 
differ from the MSW term. The reason for the latter was partly 
explained in Sec. \ref{Sec_Eff_Ev_HES}. There remains the 
disagreement of the general derivative form and the dipole derivative
form of this Hamiltonian in \cite{Kovner:2005uw}. It is the 
purpose of this section to show that one should not compare these 
Hamiltonians among themselves as well as comparing them with the 
MSW term directly because the specific form of the Hamiltonian itself 
is not unique hence the scheme dependence (see below). For this reason 
it only makes sense to compare physical quantities such as the 
action of these Hamiltonians on the target wave function. Those
are the quantities that should be in agreement independent of the
form of the Hamiltonian. 

From the discussion in Sec. 3 it is clear that our general form of 
$W_Y$, evolved from a parent dipole with transverse coordinate
$(\bx,\by)$, is 
\be W_Y(\bx,\by) = \sum^\infty_{N=1} \int d\Gamma_N P_N(Y)
    \prod^N_{i=1} \int_{\bx_{i-1}, \bx_i} R(\bx_{i-1}, \bx_i)\,\d[\r] 
\label{eq:W_Y} 
\ee
where $R(\bx,\by)$ is as given in \eref{eq:dual-dipole-2} while 
$P_N(Y)$ gives the probability to find a system of $N$ ``dipoles''
at transverse coordinates $(\bx_0,\bx_1)$, $(\bx_1,\bx_2)$, $\dots$
$(\bx_{N-1},\bx_N)$ with the first coordinate of the pair 
$(\bx_{i-1},\bx_i)$ denotes a ``quark'' and the second an ``antiquark''
with $\bx_0=\bx$ and $\bx_N=\by$, matching onto the parent dipole.
One can write the evolution of $W_Y$ as 
\be \frac{\del W_Y}{\del Y} = \c \, W_Y \;. 
\label{eq:chiW}
\ee
In \eref{JIMWLK-dual} we have taken 
\be \c = -\frac{1}{2} \int_{\bm{x},\bm{y}}\, \r^a(\bm{x})\, 
          \eta^{ab}(\bm{x},\bm{y}) [-i\,\d/\d \r] \, \r^b(\bm{y})
\label{eq:chi} 
\ee
with $\eta$ given in \eref{eta-r}. While this is a natural choice for 
$\c$ it is not unique. Indeed the choice made in Eq. (2.33) of 
\cite{Kovner:2005uw} is not the same as we have taken although it is an 
equally good choice. The product $\c\, W_Y$ in \eref{eq:chiW} is unique,
however there are different $\c$'s which acting on $W_Y$, give the
same result. 

The ambiguity in $\c$ is a sort of scheme dependence and can be 
illustrated by considering the action of $\rho^a(u^+, \bu)$ on
$U^\dagger_\bx$. It is then straightforward to verify that 
\be \rho^a(u^+, \bu)\, U^\dagger_\bx \, \d[\r]
  = \{ \tilde U_\bu (u^{'+},u^+)\}^{ab}\, \rho^b(u^{'+},\bu) \,
    U^\dagger_\bx \, \d[\r]  
\ee
where 
\be \tilde U_\bu(u^{'+}, u^+) 
   = \overline{P} \exp \Big (-g \int^{u^{'+}}_{u^+} d x^+ T^a 
                  \d/\d \r^a(x^+, \bu) \Big )  
\ee 
and where $\overline{P}$ indicates that larger $x^+$-components of the
integral are put further to the right. However it is not true that  
\be \r^a(u^+,\bu) = \tilde U^{ab}_\bu (u^{'+},u^+) \, \r^b(u^{'+},\bu) \;.
\label{eq:rho_u+}
\ee 
The fact that \eref{eq:rho_u+} is not generally correct can easily be
checked by considering the commutator of the operator, $\d^c_\bx$, 
defined earlier in \eref{eq:delta-def} (we repeat the definition here 
for convenience)   
\be \d^c_\bx = g\,\int^\infty_{-\infty} dx^+ \frac{\d}{\d \r^c(x^+,\bx)} \,, 
\ee
on the left and right hand sides of \eref{eq:rho_u+}. Clearly
\be [\d^c_\bx, \r^a(u^+,\bu)] = g\,\d_{ac}\, \d^2(\bx-\bu) 
\label{eq:[or]}
\ee
while 
\be [\d^c_\bx, \tilde U^{ab}_\bu (u^{'+},u^+) \r^b(u^+,\bu)] 
   = g\,\d^2(\bx-\bu) \, \tilde U^{ac}_\bu (u^{'+}, u^+) 
\label{eq:[our]}
\ee
and the right hand sides of \eref{eq:[or]} and \eref{eq:[our]} are
not the same. 

To see this scheme dependence more explicitly we now evaluate $\c$
through terms including four factors of $\d/\d\r$, the same level of
accuracy as kept in \cite{Mueller:2005ut} and in \cite{Kovner:2005uw}.
From \eref{eta-r}, \eref{eq:chiW} and \eref{eq:chi}
\be \c   = -\frac{1}{2\pi} \int_{\bx,\by,\bz} {\cal K}(\bx,\by,\bz)
            \r^a(\bx)\, M_{ab}\, \r^b(\by) 
\label{eq:chi-2} 
\ee
where 
\be M_{ab} = \{1+\tilde U^\dagger_{\bm{x}} \tilde U_{\bm{y}} 
                -\tilde U^\dagger_{\bm{x}} \tilde U_{\bm{z}} 
                -\tilde U^\dagger_{\bm{z}} \tilde U_{\bm{y}}\}^{ab} \;.
\label{eq:Mab}
\ee
To simplify the discussion a little we suppose that $\c\, W_Y$ 
will be used to evaluate the scattering of two separate dipoles on 
the target. In that case $M_{ab}$ can be expressed in terms of 
the operators, $\d^a_\bx$, defined in \eref{eq:delta-def},   
while in the general case the $x^+$-dependence of derivatives with 
respect to $\r$ must be kept in $M_{ab}$. Then it is straightforward
to get
\bea M^{(2)}_{ab} 
     \fx &=&\fx -\{T^c T^d \}^{ab}\, (\d^c_\bx-\d^c_\bz) \,
                                     (\d^d_\by-\d^d_\bz)
\label{eq:m2}                                                  \\
     M^{(3)}_{ab} 
     \fx &=&\fx -\frac{1}{2} \{T^c T^d T^e\}^{ab}\, 
           \Big [ \d^c_\bx (\d^d_\bx-\d^d_\by) \d^e_\by
                 -\d^c_\bx (\d^d_\bx-\d^d_\bz) \d^e_\bz
                 -\d^c_\bz (\d^d_\bz-\d^d_\by) \d^e_\by
           \Big ]    
\label{eq:m3}
\eea
\bea M^{(4)}_{ab} 
     \fx &=&\fx -\frac{1}{12} \{T^c T^d T^e T^f\}^{ab}\, \Big [ 
      \d^c_\bz \d^d_\bz \d^e_\bz \d^f_\bz
   -2 \d^c_\bx \d^d_\bz \d^e_\bz \d^f_\bz 
   -2 \d^c_\bz \d^d_\bz \d^e_\bz \d^f_\by 
   +3 \d^c_\bx \d^d_\bx \d^e_\bz \d^f_\bz             \nonumber \\ 
     \fx & & \fx \hspace{3cm} 
   +3 \d^c_\bz \d^d_\bz \d^e_\by \d^f_\by 
   -2 \d^c_\bx \d^d_\bx \d^e_\bx \d^f_\bz 
   -2 \d^c_\bz \d^d_\by \d^e_\by \d^f_\by 
   +2 \d^c_\bx \d^d_\bx \d^e_\bx \d^f_\by             \nonumber \\ 
     \fx & & \fx \hspace{3cm} 
   +2 \d^c_\bx \d^d_\by \d^e_\by \d^f_\by 
   -3 \d^c_\bx \d^d_\bx \d^e_\by \d^f_\by  \Big ] \;. 
\label{eq:m4}
\eea 

A casual glance at \eref{eq:m4} reveals that this is not the same
expression as that given by Eq. (2.38) in \cite{Kovner:2005uw}.
In particular \eref{eq:m4} has terms which have no $\d_\bz$ factors
while the formula in \cite{Kovner:2005uw} has at least one factor
of $\d_\bz$ in each of its terms. As we have no reason to expect 
agreement when comparing comparable terms in $\c$ itself, the
lack of an agreement is not surprising. 

We now turn to the product $\c\, W_Y$ which is not scheme dependent,
and we shall see how the factors in \eref{eq:m2}, \eref{eq:m3}
and \eref{eq:m4} organize thermselves into a simple expression when
the appropriate factors in $W_Y$ are included. To make the calculation
easier we take $W_Y$ as 
\be W^{(0)}_Y = R(\bx,\by)\, \d[\r]  \;.
\ee 
Because of the structure of \eref{eq:W_Y}, this is no essential 
limitation. Then letting 
${\cal K}(\bx,\by,\bz) \rightarrow -\frac{1}{2} {\cal M}(\bx,\by,\bz)$, 
\be \c\, W^{(0)}_Y = -\frac{g^2}{2\pi N_c} \int_\bz 
    {\cal M} (\bx,\by,\bz) \, M_{ab} N_{ab} \, \d[\r]  
\label{eq:chiW0}
\ee
and 
\be N_{ab} = \tr (t^a U^\dagger_\bx U_\by t^b)  \;. 
\ee
It is straightforward to evaluate $N_{ab}$ at the lowest few orders 
in the number of $\r$-derivatives. The results are  
\bea N^{(0)}_{ab} \fx &=& \fx \frac{1}{2} \d_{ab}                   \\ 
     N^{(1)}_{ab} \fx &=& \fx \tr (t^b t^a t^c) (\d^c_\bx-\d^c_\by) \\ 
     N^{(2)}_{ab} \fx &=& \fx \frac{1}{2} \tr (t^b t^a t^c t^d) \, 
                          (   \d^c_\bx \d^d_\bx+\d^c_\by \d^d_\by
                           -2 \d^c_\bx \d^d_\by )  
\eea
where, as for $M_{ab}$, the superscript denotes the order of the
derivatives in $\r$. In the product of $M_{ab} N_{ab}$ appearing in
\eref{eq:chiW0} the term $M^{(2)} N^{(0)}$ gives the usual BFKL 
evolution 
\be
M_{ab}^{(2)}N^{ab\ (0)}\,=\,-\frac{N_c}2(\delta^c_{\bm{x}}-\delta^c_{\bm{z}})
(\delta^c_{\bm{y}}-\delta^c_{\bm{z}}) \,, 
\ee 
$M_{ab}^{(3)}N^{ab\ (0)}$ vanishes and 
$M^{(2)} N^{(1)}$ gives the BFKL evolution of the odderon 
\be
M_{ab}^{(2)}N^{ab\ (1)}\,=\, -\frac{N_c}8\ d^{abc}\
\Big(\delta^a_{\bm{x}}\delta^b_{\bm{y}}(\delta^c_{\bm{x}}-\delta^c_{\bm{y}})
-\delta^a_{\bm{x}}\delta^b_{\bm{z}}(\delta^c_{\bm{x}}-\delta^c_{\bm{z}})
-\delta^a_{\bm{z}}\delta^b_{\bm{y}}(\delta^c_{\bm{z}}-\delta^c_{\bm{y}})\Big)
\ee
while the combination 
\be M^{(4)}_{ab} N^{(0)}_{ab}+M^{(3)}_{ab} N^{(1)}_{ab}
   +M^{(2)}_{ab} N^{(2)}_{ab} 
\ee
gives the four derivative terms which are our main concern here. 

Using, at large $N_c$, 
\bea \tr (t^b t^a t^c t^d) 
         &=&      \frac{1}{4N_c} (\d_{ab}\d_{cd}+\d_{bd}\d_{ac}) \\ 
     (T^e T^f)_{ab}\tr(t^b t^a t^c t^d)
         &=&      \frac14\delta_{cd}\delta_{ef}
                 +\frac18\delta_{ce}\delta_{df} \\ 
     \tr (T^c T^d T^e T^f) 
         &=&     \d_{cd}\d_{ef}+\d_{cf}\d_{de}+\frac{1}{2} \d_{ce}\d_{df}
                                                                 \\ 
     (T^c T^d T^e)_{ab} \tr (t^b t^a t^f)
         &=&     -\frac{1}{4} \tr (T^c T^d T^e T^f)  
\eea 
we find 
\bea 
M_{ab}^{(4)}N^{ab\ (0)}\fx &=& \fx 
  -\frac{5}{48}(\delta_{\bm{z}}.\delta_{\bm{z}})^2
  +\frac{5}{24}\delta_{\bm{z}}.\delta_{\bm{z}}
   (\delta_{\bm{z}}.\delta_{\bm{x}}+\delta_{\bm{z}}.\delta_{\bm{y}})
  -\frac{1}{8}\delta_{\bm{z}}.\delta_{\bm{z}}
   (\delta_{\bm{x}}.\delta_{\bm{x}}+\delta_{\bm{y}}.\delta_{\bm{y}})
      \nonumber \\     \fx & & \fx    
  -\frac{3}{16} ((\delta_{\bm{x}}.\delta_{\bm{z}})^2
                  +(\delta_{\bm{y}}.\delta_{\bm{z}})^2)
  +\frac{5}{24}
   (\delta_{\bm{x}}.\delta_{\bm{x}}\delta_{\bm{x}}.\delta_{\bm{z}}
  +\delta_{\bm{y}}.\delta_{\bm{y}}\delta_{\bm{y}}.\delta_{\bm{z}})
      \nonumber \\     \fx & & \fx     
  -\frac{5}{24}\delta_{\bm{x}}.\delta_{\bm{y}}
   (\delta_{\bm{x}}.\delta_{\bm{x}}+\delta_{\bm{y}}.\delta_{\bm{y}})
  +\frac{1}{8}\delta_{\bm{x}}.\delta_{\bm{x}}\delta_{\bm{y}}.\delta_{\bm{y}}
  +\frac{3}{16}(\delta_{\bm{x}}.\delta_{\bm{y}})^2 \;, 
\eea
%
%
\bea
M_{ab}^{(3)}N^{ab\ (1)} \fx &=& \fx \phantom{+} 
   \frac{1}{8}\delta_{\bm{z}}.\delta_{\bm{z}}
  (\delta_{\bm{x}}.\delta_{\bm{x}}+\delta_{\bm{y}}.\delta_{\bm{y}}-2\delta_{\bm{x}}.\delta_{\bm{y}})
  +\frac{1}{8}(\delta_{\bm{x}}.\delta_{\bm{x}}\delta_{\bm{y}}.\delta_{\bm{z}}
  +\delta_{\bm{y}}.\delta_{\bm{y}}\delta_{\bm{x}}.\delta_{\bm{z}})
    \nonumber\\         \fx & & \fx    
  +\frac{3}{16}(\delta_{\bm{x}}.\delta_{\bm{z}}-\delta_{\bm{y}}.\delta_{\bm{z}})^2
  -\frac{5}{16}(\delta_{\bm{x}}.\delta_{\bm{x}}\delta_{\bm{x}}.\delta_{\bm{z}}
  +\delta_{\bm{y}}.\delta_{\bm{y}}\delta_{\bm{y}}.\delta_{\bm{z}})
    \nonumber\\         \fx & & \fx    
  +\frac{3}{16}\delta_{\bm{x}}.\delta_{\bm{y}}
   (\delta_{\bm{x}}.\delta_{\bm{z}}+\delta_{\bm{y}}.\delta_{\bm{z}})
  +\frac{5}{16}\delta_{\bm{x}}.\delta_{\bm{y}}
   (\delta_{\bm{x}}.\delta_{\bm{x}}+\delta_{\bm{y}}.\delta_{\bm{y}})
    \nonumber\\         \fx & & \fx     
  -\frac{1}{4}\delta_{\bm{x}}.\delta_{\bm{x}}\delta_{\bm{y}}.\delta_{\bm{y}}
  -\frac{3}{8}(\delta_{\bm{x}}.\delta_{\bm{y}})^2 \; ,
\eea
%
%
and  
\bea
M_{ab}^{(2)}N^{ab\ (2)} \fx &=& \fx 
  -\frac{1}{8}\delta_{\bm{z}}.\delta_{\bm{z}}
   (\delta_{\bm{x}}.\delta_{\bm{x}}+\delta_{\bm{y}}.\delta_{\bm{y}}-2\delta_{\bm{x}}.\delta_{\bm{y}})
  -\frac{1}{16}(\delta_{\bm{x}}.\delta_{\bm{z}}-\delta_{\bm{y}}.\delta_{\bm{z}})^2
   \nonumber\\          \fx & & \fx 
  +\frac{3}{16}(\delta_{\bm{x}}.\delta_{\bm{x}}\delta_{\bm{x}}.\delta_{\bm{z}}
  +\delta_{\bm{y}}.\delta_{\bm{y}}\delta_{\bm{y}}.\delta_{\bm{z}})
  -\frac{3}{16}\delta_{\bm{x}}.\delta_{\bm{y}}(\delta_{\bm{x}}.\delta_{\bm{z}}+
   \delta_{\bm{y}}.\delta_{\bm{z}})
   \nonumber\\          \fx & & \fx 
  -\frac{3}{16}\delta_{\bm{x}}.\delta_{\bm{y}}
   (\delta_{\bm{x}}.\delta_{\bm{x}}+\delta_{\bm{y}}.\delta_{\bm{y}})
  +\frac{1}{8}\delta_{\bm{x}}.\delta_{\bm{x}}\delta_{\bm{y}}.\delta_{\bm{y}}
  +\frac{1}{4}(\delta_{\bm{x}}.\delta_{\bm{y}})^2\ .
\eea

Putting everything together, the result is
\bea              \fx & & \fx 
M_{ab}^{(4)}N^{ab\ (0)} +M_{ab}^{(3)}N^{ab\ (1)} +M_{ab}^{(2)}N^{ab\ (2)} 
     \nonumber \\ \fx &=& \fx \frac{1}{48}
\Big \{10\delta_{\bm{z}}.\delta_{\bm{z}}
   (\delta_{\bm{z}}.\delta_{\bm{x}}+\delta_{\bm{z}}.\delta_{\bm{y}})
  -5(\delta_{\bm{z}}.\delta_{\bm{z}})^2-6\delta_{\bm{z}}.\delta_{\bm{z}}
    (\delta_{\bm{x}}.\delta_{\bm{x}}+\delta_{\bm{y}}.\delta_{\bm{y}})
     \nonumber \\ \fx & & \fx \hspace{0.7cm} 
  -3( (\delta_{\bm{x}}.\delta_{\bm{z}})^2+(\delta_{\bm{y}}.\delta_{\bm{z}})^2
     -(\delta_{\bm{x}}.\delta_{\bm{y}})^2)
  +4(\delta_{\bm{x}}.\delta_{\bm{x}}\delta_{\bm{x}}.\delta_{\bm{z}}
  +\delta_{\bm{y}}.\delta_{\bm{y}}\delta_{\bm{y}}.\delta_{\bm{z}})
     \nonumber \\ \fx & & \fx \hspace{0.7cm} 
  +6 (\delta_{\bm{x}}.\delta_{\bm{x}}\delta_{\bm{y}}.\delta_{\bm{z}}
  + \delta_{\bm{y}}.\delta_{\bm{y}}\delta_{\bm{x}}.\delta_{\bm{z}})
  -12\delta_{\bm{x}}.\delta_{\bm{z}}\delta_{\bm{y}}.\delta_{\bm{z}}
  -4\delta_{\bm{x}}.\delta_{\bm{y}} 
   (\delta_{\bm{x}}.\delta_{\bm{x}}+\delta_{\bm{y}}.\delta_{\bm{y}}) 
   \Big \} 
    \nonumber  \\ 
\eea 

Referring back to \eref{eq:chiW0} one now finds the equality 
\be \c\, W^{(0)}_Y = -\frac{g^2 N_c}{2\pi} \int_\bz {\cal M}(\bx,\by,\bz)
                     \Big [R(\bx,\by)-R(\bx,\bz) R(\bz,\by) \Big ]_{(4)} 
                     \,\d[\r] 
\label{eq:chiW0-4thO} 
\ee 
where the subscript $(4)$ in \eref{eq:chiW0-4thO} indicates that 
the term in brackets is to be evaluated to fourth order in 
$\r$-derivatives. We note for completeness that 
\be R^{(0)}_{(2)} (\bx,\by) = \frac{1}{4 N_c} (\d_\bx-\d_\by)^2 
\label{eq:R2}
\ee 
\be
R^{(0)}_{(3)}(\bx,\by)= \frac{1}{6 N_c}\, \tr (t^a t^b t^c) 
(\d^a_\bx\d^b_\bx\d^c_\bx-3\d^a_\bx\d^b_\bx\d^c_\by
+3\d^a_\bx\d^b_\by\d^c_\by-\d^a_\by\d^b_\by\d^c_\by) 
\ee 
and 
\be R^{(0)}_{(4)} (\bx,\by) = \frac{1}{96 N_c^2} 
    [ 3 (\d_\bx-\d_\by)^2 (\d_\bx-\d_\by)^2 -\d_\bx^2 \d_\bx^2
     -\d_\by^2 \d_\by^2 -6 (\d_\bx .\d_\by)^2 
     +4 \d_\bx \cdot \d_\by (\d_\bx^2+\d_\by^2) ] \;.
\ee

It is now clear that the fourth order terms in \eref{eq:chiW0}  
agrees with the general expression \eref{final-dual-kov} derived 
earlier. It is also clear that the simple and elegant form embodied in
\eref{final-dual-kov} and \eref{eq:chiW0-4thO} is not embodied 
in $\c$ alone but is rather a property of the product $\c\,W_Y$.


\section*{Acknowledgements}

A.~Sh. acknowledges financial support by the Deutsche Forschungsgemeinschaft 
under contract Sh 92/1-1. C.M. wishes to thank the Physics Department 
of Columbia University for hospitality during his visit.

%
%
  
%
%
\end{document}